\documentclass[superscriptaddress,
 amsmath,
 amssymb,
 aps,
 prl,
 reprint
]{revtex4-2}

\usepackage{graphicx}
\usepackage{bm}
\usepackage{hyperref}

\usepackage{xcolor}
\hypersetup{
    colorlinks,
    linkcolor={black!50!black},
    citecolor={black!50!black},
    urlcolor={black!80!black}
}

\begin{document}

\title{Phonon Linewidths in Twisted Bilayer Graphene near Magic Angle}

 \author{Shinjan Mandal}
 \affiliation{Center for Condensed Matter Theory, Department of Physics, Indian Institute of Science, Bangalore}
 \author{Indrajit Maity}
 \affiliation{Departments of Materials and the Thomas
Young Centre for Theory and Simulation of Materials,
Imperial College London, South Kensington Campus}
 \author{H R Krishnamurthy}
 \affiliation{Center for Condensed Matter Theory, Department of Physics, Indian Institute of Science, Bangalore}
 \affiliation{International Centre for Theoretical Sciences, Tata Institute of Fundamental Research, Bangalore}
 \author{Manish Jain}
 \email{mjain@iisc.ac.in}
 \affiliation{Center for Condensed Matter Theory, Department of Physics, Indian Institute of Science, Bangalore}
 \date{\today}
 
\begin{abstract}
We present a computational study of the phonon linewidths in twisted bilayer graphene arising from electron-phonon interactions and anharmonic effects.
The electronic structure is calculated using distance-dependent transfer integrals based on the atomistic Slater-Koster tight-binding formalism including electron-electron interactions treated at the Hartree level, and the phonons are calculated using classical force fields. These ingredients are used to calculate the phonon linewidths arising from electron-phonon interactions. Furthermore, anharmonic effects on the linewidths are computed using the mode-projected velocity autocorrelation function obtained from classical molecular dynamics. We predict a moiré potential induced splitting of this mode, which arises due to contributions from high symmetry stacking regions. Our findings show that both electron-phonon and anharmonic effects have a significant impact on the linewidth of the Raman active G mode near the magic angle. 
\end{abstract}

\keywords{Electron-phonon interactions, twisted bilayer graphene}

\maketitle

The exceptional electronic and vibrational properties of twisted bilayer graphene (TBG) have fueled significant research interest.
The formation of large period moiré structures containing several thousand atoms, arising from a small rotational misalignment between the two graphene sheets in these systems leads to unique properties that are distinct from those of aligned bilayer graphene.
These include the emergence of flat bands \cite{cao2018unconventional, lisi2021observation, utama2021visualization, mesple2021heterostrain}, superconductivity near the magic angle of $\sim 1.1^{\circ}$ twist \cite{cao2018unconventional, yankowitz2019tuning, lu2019superconductors, oh2021evidence}, ferromagnetism \cite{sharpe2019emergent, tschirhart2021imaging}, correlated electronic behaviour \cite{cao2018correlated, lu2019superconductors, saito2020independent}, and a variety of other exotic electronic properties \cite{kerelsky2019maximized, shallcross2008quantum, shallcross2010electronic, nam2017lattice, ghawri2022breakdown, nuckolls2020strongly, polshyn2019large, cao2020strange}. 
At small angles, a sample of TBG under experimental conditions also undergoes significant lattice relaxation \cite{popov2011commensurate, alden2013strain, yoo2019atomic, carr2018relaxation}. 
The electronic and phononic properties of TBG are significantly influenced by these lattice deformations \cite{uchida2014atomic,huang2018topologically, lucignano2019crucial,koshino2019moire, maity2020phonons}.

A key debate in the scientific discourse on TBG revolves around the relative importance of electron-phonon (el-ph) and electron-electron interactions in determining the material's electronic properties, \cite{wu2018theory, koshino2018maximally,liu2018chiral,wu2019phonon,10.1063/pt.jvsd.yhyd} especially at the magic angle, where exotic properties are observed \cite{sarma2020electron}.
While evidences of correlated electron states have been observed, a comprehensive understanding of the effect of electron-phonon coupling in TBG remains a topic of intense interest in the field.

One experimentally accessible approach to investigate the strength of el-ph interactions is to measure the phonon lifetimes in Raman spectroscopy.
The phonon linewidth, which is the inverse of the lifetime, quantifies the decay rate of phonons resulting from various scattering processes caused by el-ph and other interactions. 
El-ph interactions also play a crucial role in determining the thermal conductivity, electrical resistivity, and mechanical properties of materials \cite{grimvall1976electron, RevModPhys.89.015003}.
Recent experiments in TBG have revealed the localization of the 
Raman active G mode 
\cite{gadelha2021localization} as well as a significant enhancement of the G mode linewidth near the magic angle \cite{gadelha2022electron}. Consequently, a theoretical computation of the phonon linewidths is of great interest to the community.

While the computation of the phonon linewidth is essential, first-principles calculations in TBG are extremely challenging due to the large number of atoms in the moir\'{e} unit cell. 
In this letter, we present a computational study of the electron-phonon and phonon-phonon contributions to the phonon linewidth in TBG.
Our approach combines an atomistic tight-binding model to compute the electronic structure, classical force fields to model the phonon spectra, and classical molecular dynamics (MD) simulations to account for anharmonic effects.
We investigate the influence of the moir\'{e} superlattice on the behaviour of the G mode by considering various twist angles in TBG.
By systematically varying the twist angle, the doping and the temperature, we explore the impact of the moir\'{e} superlattice on the linewidth and provide insights into the electron-phonon and phonon-phonon scattering mechanisms near the magic angle.

Assuming Matthiessen's rule, the phonon linewidth $(\Gamma)$ is the sum of the contributions from the electron-phonon interactions $(\Gamma_{\text{el-ph}})$, anharmonic effects $(\Gamma_{\text{anhm}})$ and disorder effects $(\Gamma_{\text{disod}})$ \cite{grimvall1976electron}. 
\begin{equation}
    \Gamma = \Gamma_{\text{el-ph}} + \Gamma_{\text{anhm}} + \Gamma_{\text{disod}}
    \label{eqn:1}
\end{equation}
In a pristine crystal, $\Gamma_{\text{disod}}$ is assumed to be negligible. 

Within the framework of many-body perturbation theory, the lifetime arising due to el-ph interaction for a specified phonon mode $(\mathbf{q}s)$ of the moiré superlattice is given by \cite{RevModPhys.89.015003,park2008electron}:
\begin{align}
    \Gamma_{\text{el-ph}}^{\mathbf{q}s} = \frac{4\pi}{\Omega_{\text{BZ}}} \sum_{m,n}\int \text{d}\mathbf{k} \  |g_{\mathbf{k},\mathbf{q}}^{mns}|^2  &(f_{\mathbf{k}}^{n} - f_{\mathbf{k+q}}^{m}) \times \nonumber \\ 
			 &\delta(\epsilon_{\mathbf{k+q}}^{m}-\epsilon_{\mathbf{k}}^{n}-\omega_{\mathbf{q}s})
    \label{eqn:Elphlwtdh}
\end{align}
where the el-ph coupling matrix element
\begin{equation}
    g_{\mathbf{k},\mathbf{q}}^{mns} = \langle \psi_{\mathbf{k+q}}^m |\nabla_{\mathbf{q}s} H |\psi_{\mathbf{k}}^n \rangle
\end{equation}
is the amplitude for the scattering of an electron from the $n^{\text{th}}$ electronic band at $\mathbf{k}$ to the $m^{\text{th}}$  band at $\mathbf{k+q}$ by interacting with the phonon mode $(\mathbf{q}s)$. The $n^{\text{th}}$ electronic band at $\mathbf{k}$ has an energy $\epsilon_{\mathbf{k}}^{n}$ and a Fermi occupation of $f_{\mathbf{k}}^{n}$, while the phonon mode $(\mathbf{q}s)$ has energy $\omega_{\mathbf{q}s}$. We describe below the procedure we use for the evaluation of $g_{\mathbf{k},\mathbf{q}}^{mns}$.
 
\begin{figure}
    \centering
    \includegraphics[width=0.48\textwidth]{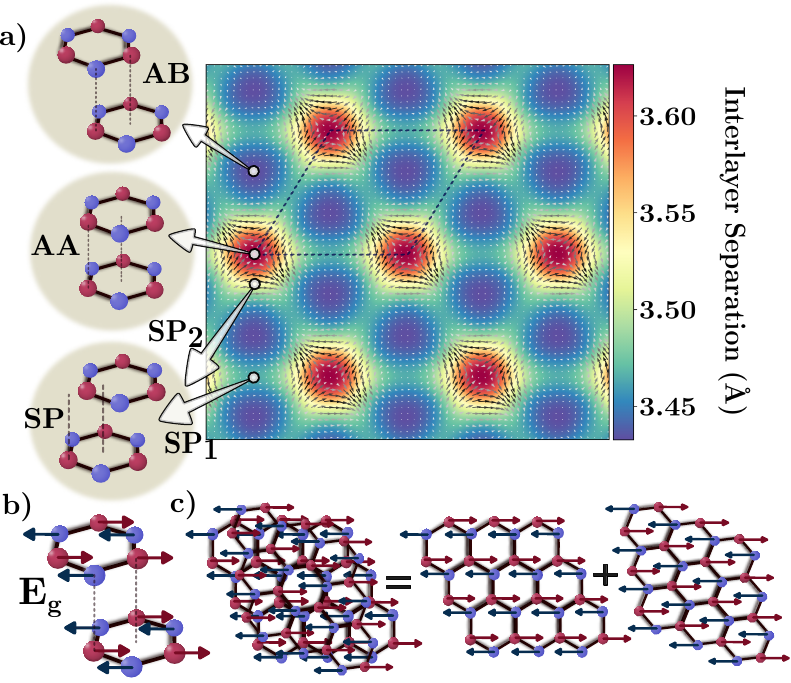}
    \caption{a) Interlayer separation of $1.05^{\circ}$ TBG including atomic relaxations. Atomic structures associated with the different stacking regions (AA, AB, and SP) regions are shown. The gradient of the interlayer separation is shown by the arrows. We can further categorize the SP regions into SP$_1$ and SP$_2$ zones by distinguishing between the SP regions which have a higher gradient.
    b) A schematic of the eigenvectors of the $\text{E}_{\text{g}}$ mode at the $\Gamma$ point in AB stacked bilayer graphene. The direction  of displacement of each atom is shown by the corresponding arrows.
    c) A schematic of the bilayer like $\text{E}_{\text{g}}$ mode at the $\Gamma$ point in twisted bilayer graphene. We assume the displacement of all the A (red dots) and B sites (blue dots) in the supercell to be identical to the corresponding displacements of the A and B atoms in the AB stacked bilayer graphene.
    }
    \label{fig:1}
\end{figure}

We generate the TBG structures using the \texttt{TWISTER} package \cite{naik2022twister}. Structural relaxations of the atoms in TBG are performed using classical interatomic potentials  implemented in the \texttt{LAMMPS} package~\cite{thompson2022lammps}. Specifically, we use the Tersoff \cite{kinaci2012thermal} potential for the intralayer interactions and the DRIP \cite{wen2018dihedral}  potential for the inter-layer interactions. 
The interlayer separation for the relaxed structure at $1.05^{\circ}$ twist is shown in Fig. (\ref{fig:1}a). To obtain the electronic structure of the relaxed structure we have used an atomistic tight-binding model with distance-dependent Slater-Koster transfer integrals \cite{slaterkoster}.
\begin{equation}
\hat{\mathbf{H}} = -\sum\nolimits_{i,j} t(\mathbf{r}_i-\mathbf{r}_j)c_i^{\dagger}c_j + \text{h.c.}	
\end{equation}
The doping-dependent changes to the electronic structure are incorporated into our tight-binding model via on-site Hartree interactions, following ref  \cite{goodwin2020hartree}: 
\begin{equation}
V_H(\mathbf{r}) = V(\theta)(\nu - \nu_0)\sum\nolimits_{j=1,2,3} \cos (\mathbf{G}_j . \mathbf{r})    
\end{equation}
Here $\nu_{0}(\theta)$ is the reference doping level where the Hartree potential is zero, $V(\theta)$ is a twist angle dependent energy parameter and $\mathbf{G}_j$ denote the three reciprocal lattice vectors of the moir\'{e} unit cell. The parameters for our calculations were obtained from \cite{goodwin2020hartree}.
We compute the force constants required to construct the dynamical matrix, $D(\mathbf{q})$, for the relaxed structures from the same classical interatomic potentials. Diagonalizing $D(\mathbf{q})$ at each $\mathbf{q}$ point we obtain
\begin{equation}
    D(\mathbf{q}) \Psi_{\mathbf{q}s}^{\text{M}}= \omega_{\mathbf{q}s}^{2}\Psi_{\mathbf{q}s}^{\text{M}} 
\end{equation}
$\Psi_{\mathbf{q}s}^{\mathbf{M}}$ is the eigenvector of the moiré phonon mode $(\mathbf{q}s)$. 
In the localized atomic orbital framework, the el-ph coupling matrix elements takes the form \cite{agapito2018ab,choi2018strong},
\begin{multline}
        g_{\mathbf{k},\mathbf{q}}^{mns} =  l_{\mathbf{q}s}\sum_{\kappa\alpha}\Psi_{\mathbf{q}s, \kappa\alpha}^{\text{M}}\sum_{pi}\frac{\partial}{\partial x_{\alpha}} t(\tau_{0}^{\kappa} - \tau_{p}^{i}) \\
	\left( e^{i\mathbf{k}\mathbf{R_p}} \phi_{\mathbf{k+q},m\kappa}^{*} \phi_{\mathbf{k},ni}  + 
	e^{-i(\mathbf{k+q})\mathbf{R_p}} \phi_{\mathbf{k+q},mi}^{*}\phi_{\mathbf{k},n\kappa}\right)
    \label{eqn:coupling}
\end{multline}
Here $\tau_{p}^{i}$ denotes the position of the $i^{th}$ atom in the moir\'{e} unit cell centered at $\mathbf{R_p}$,  $\phi_{\mathbf{k},ni}$ is the component of the electronic eigenvetor for the ${n^{\text{th}}}$ band with the wave vector $\mathbf{k}$, in the local orbital basis, and $l_{\mathbf{q}s}=\sqrt{\hbar/2M_c\omega_{\mathbf{q}s}}$ 
is the characteristic length scale associated with each phonon mode, with $M_c$ denoting the mass of a carbon atom.
Using eqn.(\ref{eqn:coupling}) and eqn.(\ref{eqn:Elphlwtdh}), we get the contribution from each phonon moir\'{e} mode to $\Gamma_{\text{el-ph}}$. 
In principle, the electron-phonon coupling matrix element would have a further contribution from the gradient of the on-site Hartree term as well. But within our framework, the effect of this term on the computed linewidths turn out to be negligible in the case of TBG. Further details are provided in the Supplementary Information.
Previous studies on twisted bilayer transition metal dichalcogenides at finite temperature found that the bandwidth changes were approximately $\sim 2-3$ meV for both the valence and conduction bands \cite{maity2023electrons}.
Anticipating a comparable temperature dependence in electronic bandwidths for TBG, we expect that thermal effects on the electronic band structure would not substantially alter the electron-phonon contribution to the linewidth of the G modes ($\sim 200$ meV), and hence have not been considered in these calculations.

\begin{figure}
    \centering
    \includegraphics[width=0.5\textwidth]{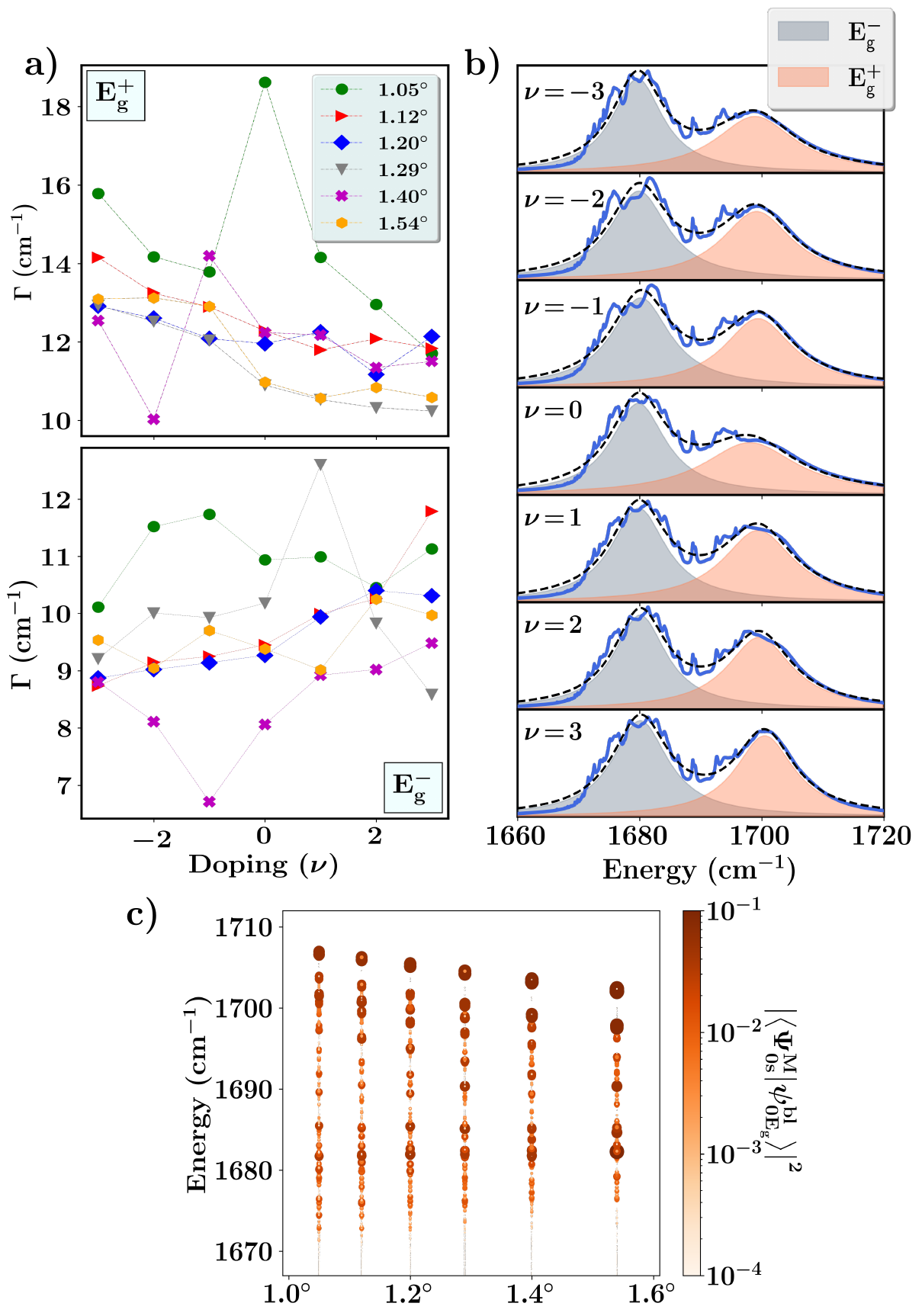}
    \caption{a) The el-ph interaction induced linewidths of the E$_\text{g}^{+}$ (top) and the E$_\text{g}^{-}$ (bottom) modes as a function of doping for twist angles near the magic angle. b) A representative fit for the el-ph spectral function at different dopings $(\nu)$ is shown for the twist angle of $1.05^{\circ}$. The fits for the other angles are available in the Supplementary Information. c) The projection of the moir\'{e} modes $(\Psi_{\mathbf{0}s}^{\text{M}})$ onto the bilayer-like $\text{E}_{\text{g}}$ modes $\psi_{\mathbf{0}\text{E}_{\text{g}}}^{\text{bl}}$ as a function of twist angle. The colours, as well as the sizes of the points, indicate the weights of the corresponding moir\'{e} modes.}
    \label{fig:2}
\end{figure}

In order to calculate the experimentally observed Raman spectra, the Raman intensities for each of the moir\'e modes should be calculated \cite{PhysRevB.88.035428,PhysRevB.108.125401,sato2012zone,saito1998raman,jorio2011raman}. However, as this would be prohibitively expensive for angles close to the
magic angle, we have chosen a different procedure.
Due to the reduced symmetry of the moir\'{e} unit cell, the moir\'{e} phonon modes cannot be labeled with the irreducible representations of the D$_{3d}$ group.
To take the effects of the relxation into account, at $\mathbf{q} = \mathbf{0}$, we can define a bilayer-like mode in the relaxed moir\'{e} unit cell by translating the in-plane G mode atomic displacements in AB stacked bilayer graphene (Fig (\ref{fig:1}b))  to the corresponding atoms in the moir\'{e} (Fig (\ref{fig:1}c)).
We denote the eigenvector for this bilayer-like G mode in the moire unit cell as $\psi_{\mathbf{0}\text{E}_{\text{g}}}^{\text{bl}}$.
The linewidth arising from each moiré phonon mode is computed by projecting the moir\'{e} unit cell phonon modes onto this bilayer-like phonon mode and constructing a spectral function as
\begin{align}
    A_{\mathbf{0}\text{E}_{\text{g}}}(\omega) = \frac{1}{\pi} \sum_{s} \frac{\big\lvert \langle \Psi_{\mathbf{0}s}^{\text{M}} \lvert \psi_{\mathbf{0}\text{E}_{\text{g}}}^{\text{bl}} \rangle \big\rvert^2 \ \Gamma_{\text{el-ph}}^{\mathbf{0}s}}{(\omega - \omega_{\mathbf{0}s})^2 + (\Gamma_{\text{el-ph}}^{\mathbf{0}s})^2}
    \label{eqn:spectral_fn}
\end{align}
Fitting a Lorentzian to this spectral function, we obtain the el-ph linewidth corresponding to the bilayer-like unit cell phonon mode.
The results are shown in Figs. \ref{fig:2}a and \ref{fig:2}b.
Previous investigations into the computation of G mode linewidths in TBG near the magic angle \cite{gadelha2022electron} employed a method similar to ours, where linewidths were calculated for both Raman active and inactive (or weakly active) modes via Eqn.  (\ref{eqn:coupling}) and the eventual linewidth of the G mode was obtained by employing weight factors that effectively filtered out the contributions from the Raman-inactive modes. 
In our approach of constructing the spectral function, we capture the G-mode-like characteristics of moiré modes, incorporating their frequencies as well. This comprehensive approach provides an improved understanding of the anticipated Raman activity across the frequency spectrum.  

Fig \ref{fig:2}c shows that there is a systematic splitting of the projected weights of the moir\'e mode onto the bilayer-like G mode with decreasing twist angle, resulting in two distributions: a high-frequency distribution (around 1700 cm$^{-1}$), which we call the E$_{\text{g}}^{+}$ distribution, and a low-frequency distribution (around 1680 cm$^{-1}$), which we call the E$_{\text{g}}^{-}$ distribution.
This splitting can be understood in the context of a nearly free phonon model, which has shown that gaps in the phonon bands of periodic structures can arise due to the effect of an underlying secondary long-period potential. The strength of this potential modulates the size of the gap \cite{van2022nearly}.
In the case of twisted bilayer graphene, the moiré superlattice creates a secondary long-period moiré potential that splits the phonon bands  \cite{PhysRevLett.109.186807}.
The increase in the split of the G mode as a function of twist near the magic angle shows that the moiré potential becomes stronger as one approaches the magic angle.

We show the linewidths of E$_{\text{g}}^{+}$  and E$_{\text{g}}^{-}$ mode of the moir\'{e} unit cell obtained as a function of doping for several twist angles around the magic angle Fig(\ref{fig:2}a). 
The linewidth of the E$_{\text{g}}^{+}$ mode decreases as doping increases (from -3 to +3), while the linewidth of the E$_{\text{g}}^{-}$ mode increases as doping increases. 
The anomalous increase in the linewidth of the E$_\text{g}^{+}$ mode at $\nu=0$ for $1.05^{\circ}$ can be explained by the joint electronic density of states (JDOS). The peak in the JDOS aligns with the energy distribution of the E$_\text{g}^{+}$ mode, which results in an increase in the number of scattering channels due to the phonon, and hence an enhanced linewidth (see Fig S9 in the Supplementary Information).  
 
In the low-angle regime, atomic relaxation leads to the formation of high-symmetry stacking regions with differential strain profiles. Previous studies have shown that this variation in the strain profile localizes the G modes \cite{gadelha2021localization}.
We have identified the stacking regions by examining the interlayer separation between the two graphene layers.
As shown in Fig (\ref{fig:1}a) the AA and the AB stacked regions are easily identified by their respective interlayer distances. 
In previous studies, the transition region between the AA and the AB was denoted as the saddle point (SP) region \cite{gadelha2021localization}. 
In our analysis we further distinguish two types of saddle point regions, SP$_1$ and SP$_2$. 
In terms of the in-plane gradient of the interlayer separation, we can see that SP$_1$, the transition region between two adjacent AB solitonic regions, has a smaller magnitude of the gradient than SP$_{2}$, the transition region between an AA and an AB solitonic regions.   
\begin{figure}
    \centering
    \includegraphics[width=0.48\textwidth]{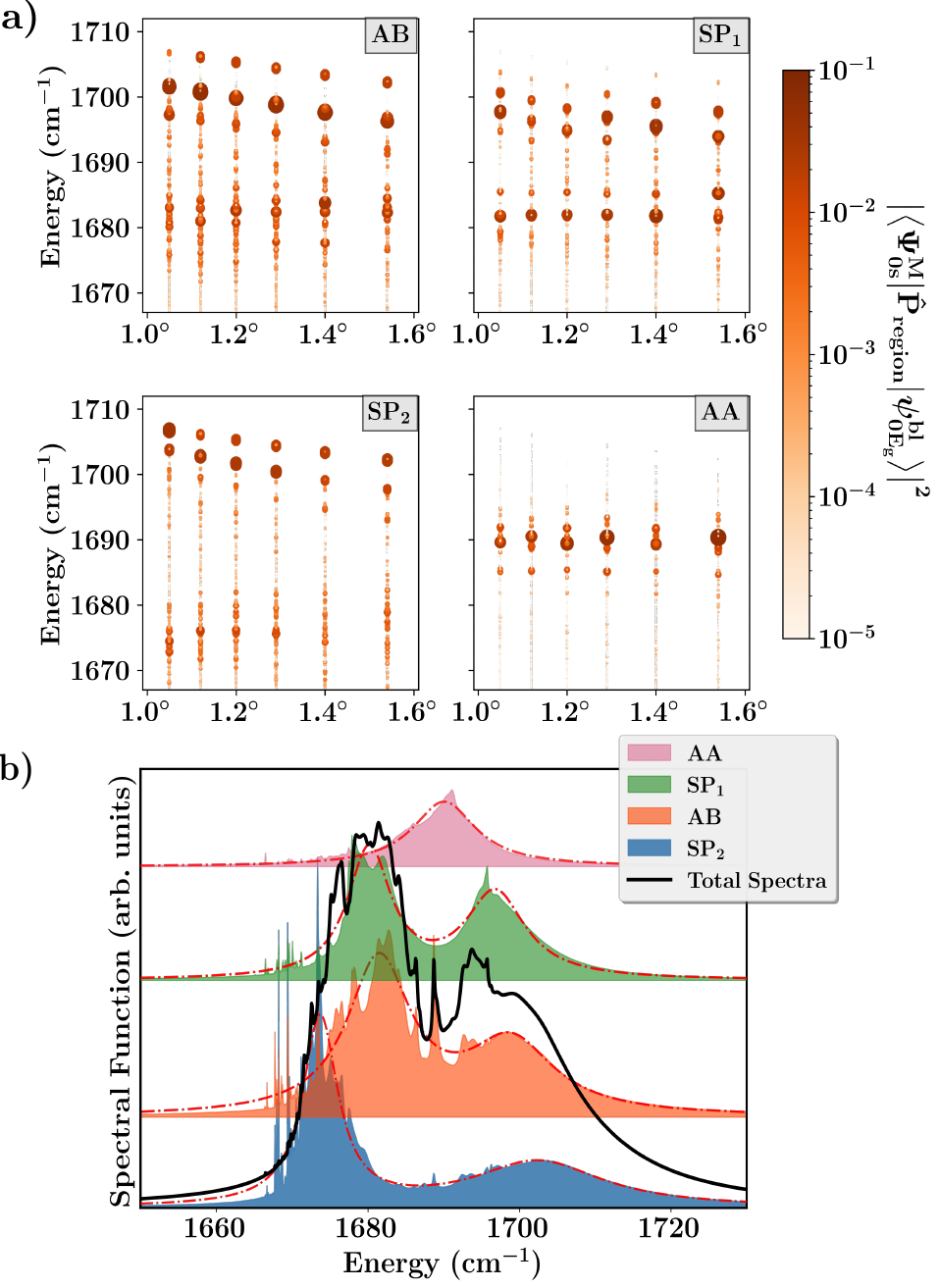}
    \caption{a) The projections of $\Psi_{\mathbf{0}s}^{\mathbf{M}}$ onto the bilayer like $\text{E}_{\text{g}}$ mode in different solitonic regions of the moir\'{e} cell. b) The projected spectral functions for $1.05^{\circ}$ TBG  at $\nu=0$ ($4^{\text{th}}$ panel in Fig \ref{fig:2}b) projected onto different regions of the moir\'{e}. The 
    red dashed lines show the best fit to the data. }
    \label{fig:3}
\end{figure}
We can project the moiré modes onto the components of $\psi_{\mathbf{0} \text{E}_{\text{g}}}^{\text{bl}}$ arising from the four stacking regions, $\lvert \langle \Psi_{\mathbf{0}s}^{\text{M}} \lvert \hat{\text{P}}_{\text{region}} \rvert \psi_{\mathbf{0} \text{E}_{\text{g}}}^{\text{bl}} \rangle \rvert^2$.
As shown in Fig. \ref{fig:3}a, we find that both high frequency and low frequency modes have some localization characteristics. 
In the AB region, the high-frequency moiré modes have significant weights, while the low-frequency modes have smaller but non-negligible weights. 
The SP$_1$ region has signatures of both high and low frequency modes, while the SP$_2$ region has significantly higher weights for the high-frequency modes.
The projection on the AA region shows that the modes at $\sim 1690$ cm$^{-1}$ are almost entirely localized there consistently for all twist angles considered in this study. 
This frequency is close to the E$_{\text{g}}$ mode frequency of untwisted bilayer graphene (1690.96 cm$^{-1}$), which is obtained using the same force fields employed in our calculations.

The identification of the localized phonon modes provides a new perspective for analyzing the spectral functions computed using eqn. (\ref{eqn:spectral_fn}).
We observe seven distinct frequency regions across the four stacking regions, where the projections onto the unit cell modes have significant weight.
For each stacking region we plot the spectral function 
\begin{equation}
    A_{\mathbf{0}\text{E}_{\text{g}}}^{\text{region}}(\omega) = \frac{1}{\pi} \sum_{s} \frac{\lvert \langle \Psi_{\mathbf{0}s}^{\text{M}} \lvert \hat{\text{P}}_{\text{region}} \rvert \psi_{\mathbf{0}\text{E}_{\text{g}}}^{\text{bl}} \rangle \rvert^2 \ \Gamma_{\text{el-ph}}^{\mathbf{0}s}}{(\omega - \omega_{\mathbf{0}s})^2 + (\Gamma_{\text{el-ph}}^{\mathbf{0}s})^2}
    \label{eqn:modified_spectral_fn}
\end{equation}
The region projected spectral function for $1.05^{\circ}$ TBG at $\nu = 0$ is shown in Fig \ref{fig:3}b.
Additional plots for different angles near the magic angle at integer fillings demonstrating similar behaviour are shown in the Supplementary Information.
Fitting the spectral function of the G mode in TBG near the magic angle with Lorentzian functions centered around the seven distinct frequencies, each of which corresponds to a unique contribution from a specific solitonic region, enables a more nuanced interpretation of the underlying physical mechanisms that govern the spectral function. 

The anharmonic contribution to the linewidth, $\Gamma_{\text{anhm}}$, is calculated from the mode projected velocity auto-correlation function (MVACF) \cite{maity2020phonons} of the bilayer like E$_{\text{g}}$ mode. 
The power spectrum constructed by taking the Fourier transform of the MVACF contains information about the line shift and linewidth of the projected phonon mode at different temperatures. 
Details about computation of the MVACF are provided in the Supplementary Information. 

Similar to the el-ph interaction contributions to the $\text{E}_{\text{g}}$ mode spectrum, the MVACF power spectrum is also expected to separate into E$_{\text{g}}^{+}$ and E$_{\text{g}}^{-}$ regions. However, unlike the el-ph interaction induced spectra, performing a spatial analysis on the MVACF spectra is non-trivial. The dynamics of the stacking regions under MD simulations make them difficult to identify. Time averaged stacking regions could potentially be defined by examining the dynamics of the solitonic regions for a finite time, but such an analysis has not been attempted in this work. 

\begin{figure}
    \centering
    \includegraphics[width=0.48\textwidth]{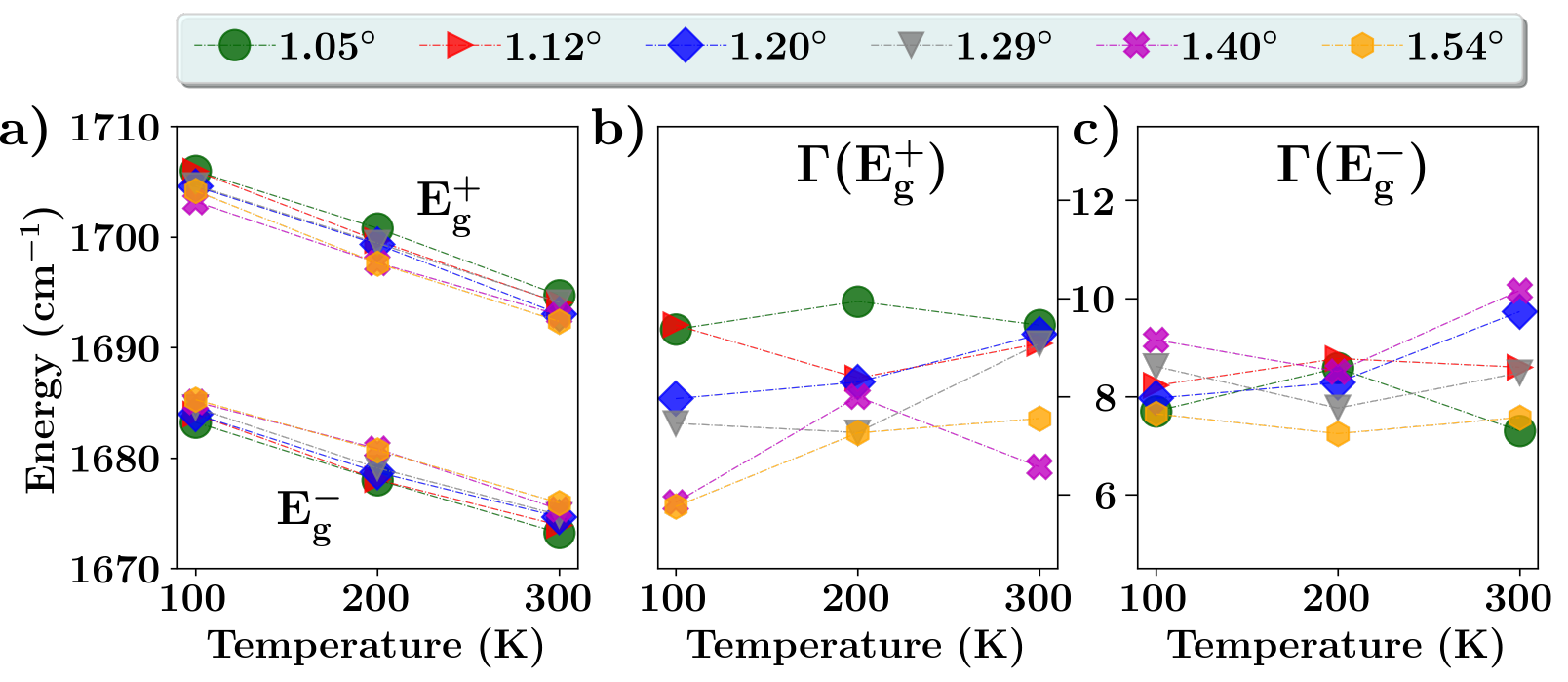}
    \caption{(a) The lineshifts arising from the anharmonic interactions for the E$_{\text{g}}^{+}$ and E$_{\text{g}}^{-}$ modes as a function of Temperature. The corresponding linewidths ($\Gamma_{\text{anhm}}$) obtained at different temperatures for each angle are also shown for E$_{\text{g}}^{+}$ (b) and E$_{\text{g}}^{-}$ (c) modes. The units for the reported linewidths are cm$^{-1}$.}
    \label{fig:4}
\end{figure}

The classical force fields used in our computations do not account for doping effects because they lack parametric dependence on the electronic interactions. Therefore, the anharmonic linewidths of the E$_{\text{g}}^{+}$ and E$_{\text{g}}^{-}$ modes are independent of doping. Consequently, the total linewidth $(\Gamma)$ obtained at each doping for a particular angle is equivalent to the el-ph linewidth $\Gamma_{\text{el-ph}}$ at the particular doping shifted by a constant amount (eqn.\ref{eqn:1})
Fitting Lorentzians to the MVACF power spectrum yields the lineshifts and linewidths of the E$_{\text{g}}^{+}$ and E$_{\text{g}}^{-}$ modes.
The lineshifts of the modes are shown in the left panel of Fig \ref{fig:4}. 
For both E$_{\text{g}}^{+}$ and E$_{\text{g}}^{-}$, the modes soften by approximately $10$ cm$^{-1}$ on increasing the temperature from $100$K to $300$K. 
The linewidths of these two modes for different twist angles are shown as a function of temperature in the center and right panels of Fig (\ref{fig:4}) respectively. 

In summary, in this theoretical investigation we have computed the contributions of el-ph and ph-ph interactions to the linewidth of the G mode phonon in TBG. 
The modulation of the moiré potential due to the twist angle introduces gaps in the phonon mode spectra at the $\Gamma$ point near the magic angle. We expect that 
this split of $\sim 20$ cm$^{-1}$ in the E$_{\text{g}}$ mode will be observable in Raman measurements. 
Our el-ph calculations show a systematic increase/decrease in the el-ph linewidth of the E$_{\text{g}}^{+}$ and E$_{\text{g}}^{-}$ modes as a function of electron doping. The ph-ph scattering calculations show a similar split of the G mode into E$_{\text{g}}^{+}$ and E$_{\text{g}}^{-}$. The linewidths arising from ph-ph scattering do not change significantly as a function of temperature.

\begin{acknowledgments}
The authors thank Kemal Atalar and Valerio Vitale for useful discussions over the course of the project and Darshit Solanki for inputs during the analysis of the data. M.J. acknowledges the National Supercomputing Mission of the Department of Science and Technology, India, and Nano Mission of the Department of Science and Technology for financial support under Grants No. DST/NSM/R\&D\_HPC\_Applications$/2021/23$ and No. DST/NM/TUE/QM-10/2019 respectively. H.R.K. acknowledges the Science and Engineering Research Board of the Department of Science and Technology, India, and the Indian National Science Academy for support under Grants No. SB/DF/$005/2017$ and No. INSA/SP/SS/$2023/$ respectively. 
H.R.K. also acknowledges support in ICTS by a grant from the Simons Foundation (677895, R.G.).
I.M. acknowledges funding from the European Union's Horizon 2020 research and innovation program under the Marie Skłodowska-Curie Grant agreement No. 101028468. 
\end{acknowledgments}

\bibliography{bibtex.bib}

\end{document}


\title{Supplementary: Phonon Linewidths in Twisted Bilayer Graphene near Magic Angle}

\author{Shinjan Mandal}
 \affiliation{Center for Condensed Matter Theory, Department of Physics, Indian Institute of Science, Bangalore}
 \author{Indrajit Maity}
 \affiliation{Departments of Materials,
Imperial College London, South Kensington Campus}
 \author{H R Krishnamurthy}
 \affiliation{Center for Condensed Matter Theory, Department of Physics, Indian Institute of Science, Bangalore}
 \affiliation{International Centre for Theoretical Sciences, Tata Institute of Fundamental Research, Bangalore}
 \author{Manish Jain}
 \email{mjain@iisc.ac.in}
 \affiliation{Center for Condensed Matter Theory, Department of Physics, Indian Institute of Science, Bangalore}

 \date{\today}

\maketitle

\setcounter{figure}{0}                       
\renewcommand\thefigure{S\arabic{figure}}   

\section{Structure Generation and Relaxation}
We generate the rigid TBG structures using the TWISTER code \cite{naik2022twister} and perform the atomic relaxation in LAMMPS \cite{thompson2022lammps} using Tersoff \cite{kinaci2012thermal} as the intralayer potential and DRIP \cite{wen2018dihedral} as the interlayer potential. The systems are relaxed upto a force tolerance of $10^{-6}$ eV/\AA.

The choice of Tersoff over other standard potentials like LCBOP or AIREBO was intentional even though the phonon frequencies of the high energy optical modes computed with this potential are grossly overestimated. The temperature dependence of the G modes computed with Tersoff shows the correct qualitative behaviour, unlike other potentials where significant hardening of the modes is observed with rising temperature, contrary to experimental observations \cite{koukaras2015phonon}. Since we have computed the contributions due to the anharmonic effects as well, ensuring the correct temperature dependence of the G-mode was crucial for our calculations.

\section{\label{tbparam}Electronic Structure}

\begin{figure}
    \centering
    \includegraphics[width=\textwidth]{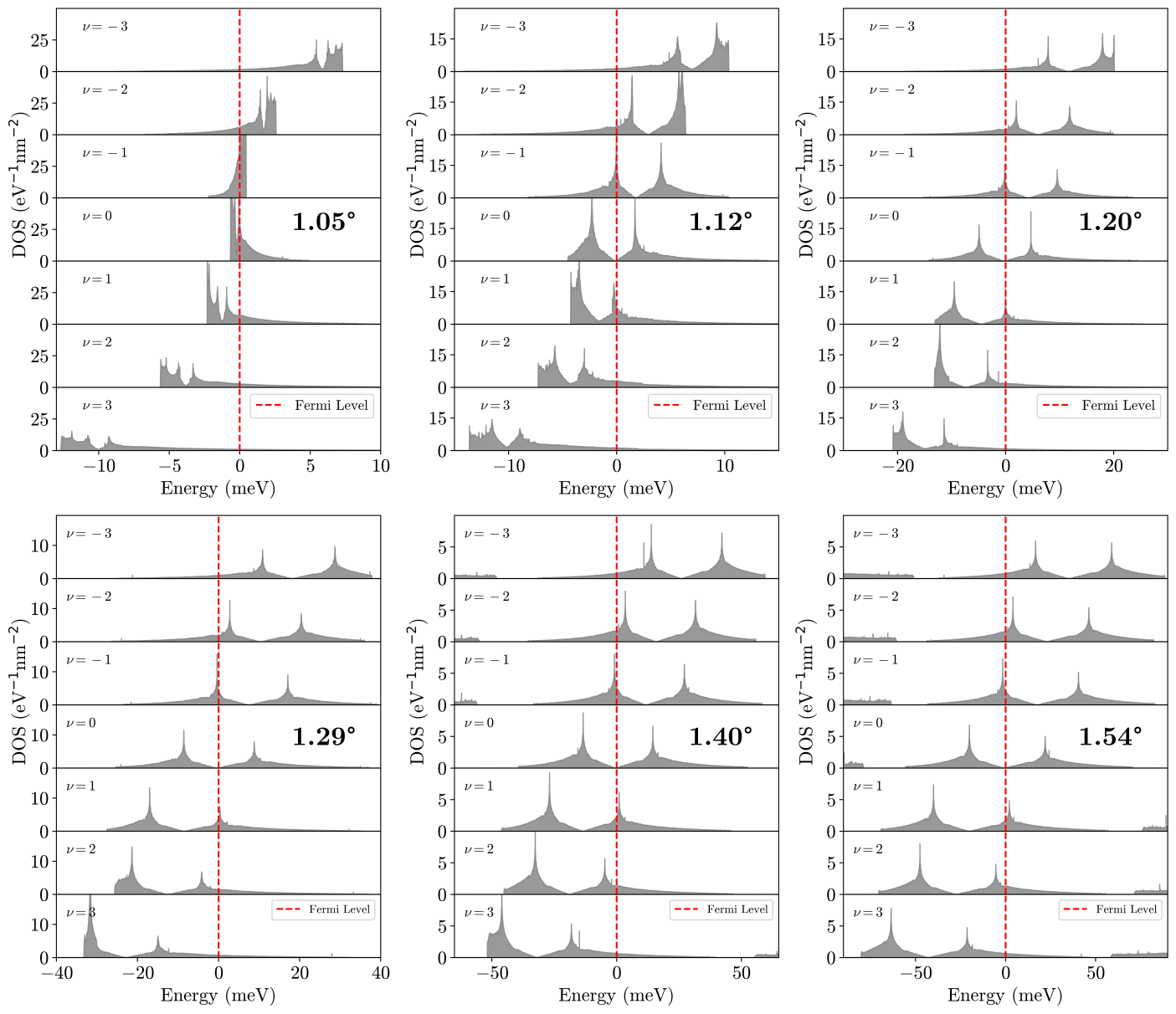}
    \caption{Density of states for the systems we have studied, as a function of the filling $\nu$. The electronic structures were computed after putting in the effects of the Hartree interactions \cite{goodwin2020hartree}}
    \label{fig:DOS_H}
\end{figure}

 The electronic hamiltonian of the system is 
\begin{equation}
\hat{\mathbf{H}} = -\sum_{i,j} t(\mathbf{r}_i-\mathbf{r}_j)c_i^{\dagger}c_j + \text{h.c.} = -\sum_{i,j} t_{ij}c_i^{\dagger}c_j + \text{h.c.}	
\end{equation}
where $\mathbf{r}_i$ denotes the real space position of the $i^\text{th}$ atom, and $c_i^{\dagger}$ and $c_i$ are the creation and annihilation operators at $\mathbf{r}_i$. We approximate the transfer integrals $t_{ij}$ using the Slater-Koster formalism \cite{slaterkoster} assuming that the overlap of the $p_z$ orbitals can be approximated as the linear combination of the $\sigma\sigma$ and $\pi\pi$ overlaps. 
Taking the local curvature of the sheets into account the transfer integral can be written as \cite{skchoicnt}:
\begin{equation}
	t_{ij}  = t_{\pi\pi}[\hat{\mathbf{n}}_i-(\hat{\mathbf{n}}_i\cdot\hat{\mathbf{r}}_{ij})\hat{\mathbf{r}}_{ij}]\cdot[\hat{\mathbf{n}}_j-(\hat{\mathbf{n}}_j\cdot\hat{\mathbf{r}}_{ij})\hat{\mathbf{r}}_{ij}]+
	t_{\sigma\sigma}[\hat{\mathbf{n}}_i\cdot\hat{\mathbf{r}}_{ij}]\cdot[\hat{\mathbf{n}}_j.\hat{\mathbf{r}}_{ij}]
	\label{transfer_integral}
\end{equation}
where $\hat{\mathbf{n}}_i$ is the unit normal at the $i^{\text{th}}$ site, and $\hat{\mathbf{r}}_{ij}$ is the unit vector joining the sites $\mathbf{r}_i$ and $\mathbf{r}_j$.
The terms $t_{\pi\pi}$ and $t_{\sigma\sigma}$ are taken as follows:
\begin{equation}
	\begin{aligned}
		&t_{\pi\pi} = t_{\pi}^0\exp\Big(-\frac{\mathbf{r}_{ij} - a_0}{\delta}\Big); \ \ \ 	t_{\pi}^0 = -2.7 \ \text{eV} \\
		&	t_{\sigma\sigma} = t_{\sigma}^0\exp\Big(-\frac{\mathbf{r}_{ij} - d_0}{\delta}\Big); \ \ \ 	t_{\sigma}^0 = 0.48 \ \text{eV}
	\end{aligned}
\end{equation}
The parameter $a_0 = 1.42 \ \text{\AA}$ is taken as the nearest neighbour distance, $d_0 = 3.35 \ \text{\AA}$ and the attenuating factor, $\delta = 0.184\sqrt{3}a_0$ is chosen so that the strength of the second nearest neighbor transfer integral is $0.1$ times that of the first neighbour.\cite{moon2012energy} \\[12pt]
The Hartree interaction is modelled as an onsite term:
\begin{equation}
    V_{H}(r) \approx V(\theta)(\nu - \nu_0(\theta))\sum_{j=1,2,3} \cos({\mathbf{G}_j\cdot \mathbf{r}}) 
\end{equation}
Here $\nu$ is the filling and $\mathbf{G}_j$ are the three reciprocal lattice vectors used to describe the out-of-plane corrugation of the TBG.
The parameters $\nu_0(\theta)$ and $V(\theta)$ have been provided in Ref.\cite{goodwin2020hartree}. The resulting electronic densities of states (DOS) for the various angles we have studied are shown in Fig. \ref{fig:DOS_H}.

\newpage
\section{Phonon Calculations}
\begin{figure}[h!]
    \centering
    \includegraphics[width=0.5\textwidth]{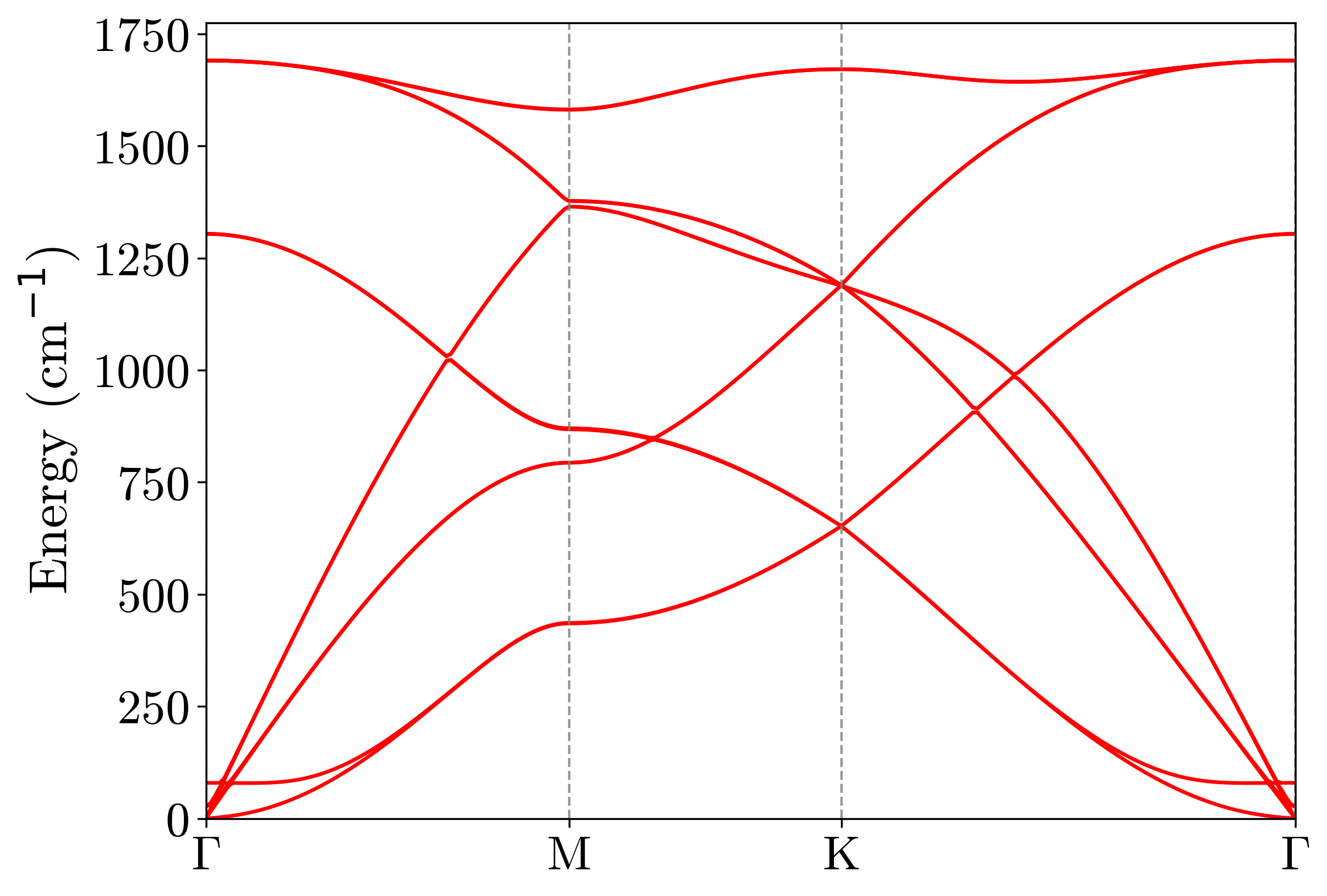}
    \caption{Phonon modes in AB stacked bilayer graphene}
    \label{fig:AB modes}
\end{figure}

The force constants were generated via a finite difference method from the classical force fields mentioned before.
\begin{equation}
	\Phi_{\alpha\beta}^{j0, j'l'}  = \frac{\partial^2 V}{\partial \mathbf{r}_{\alpha}^{j0} \partial \mathbf{r}_{\beta}^{j'l'}}
\end{equation}
where $\mathbf{r}_{\alpha}^{jl}$ denotes component of the position of the $j^{\text{th}}$ atom in the $l^{\text{th}}$ unit cell in the Cartesian direction $\alpha$. The dynamical matrix is given by 
\begin{equation}
    	D_{\alpha\beta}^{jj'}(\mathbf{q}) = \frac{1}{\sqrt{m_j m_{j'}}}
		\sum_{l'}\Phi_{\alpha\beta}^{j0, j'l'}
		e^{i\mathbf{q}\cdot(\mathbf{r}_{j'l'}-\mathbf{r}_{j0})}
\end{equation}
To obtain the moir\'{e} phonon spectrum at each $\mathbf{q}$ point, we need to solve the eigenvalue equation
\begin{equation}
    D(\mathbf{q}) \Psi_{\mathbf{q}\nu}^{\mathbf{M}} = \omega_{\mathbf{q}\nu}^{2}\Psi_{\mathbf{q}\nu}^{\mathbf{M}} 
\end{equation} 
The AB stacked bilayer graphene phonon modes obtained using the  potentials are shown in Fig \ref{fig:AB modes}.

\subsection{\textbf{Identification of the modes}}
\begin{figure}[h!]
    \centering
    \includegraphics[width=0.85\textwidth]{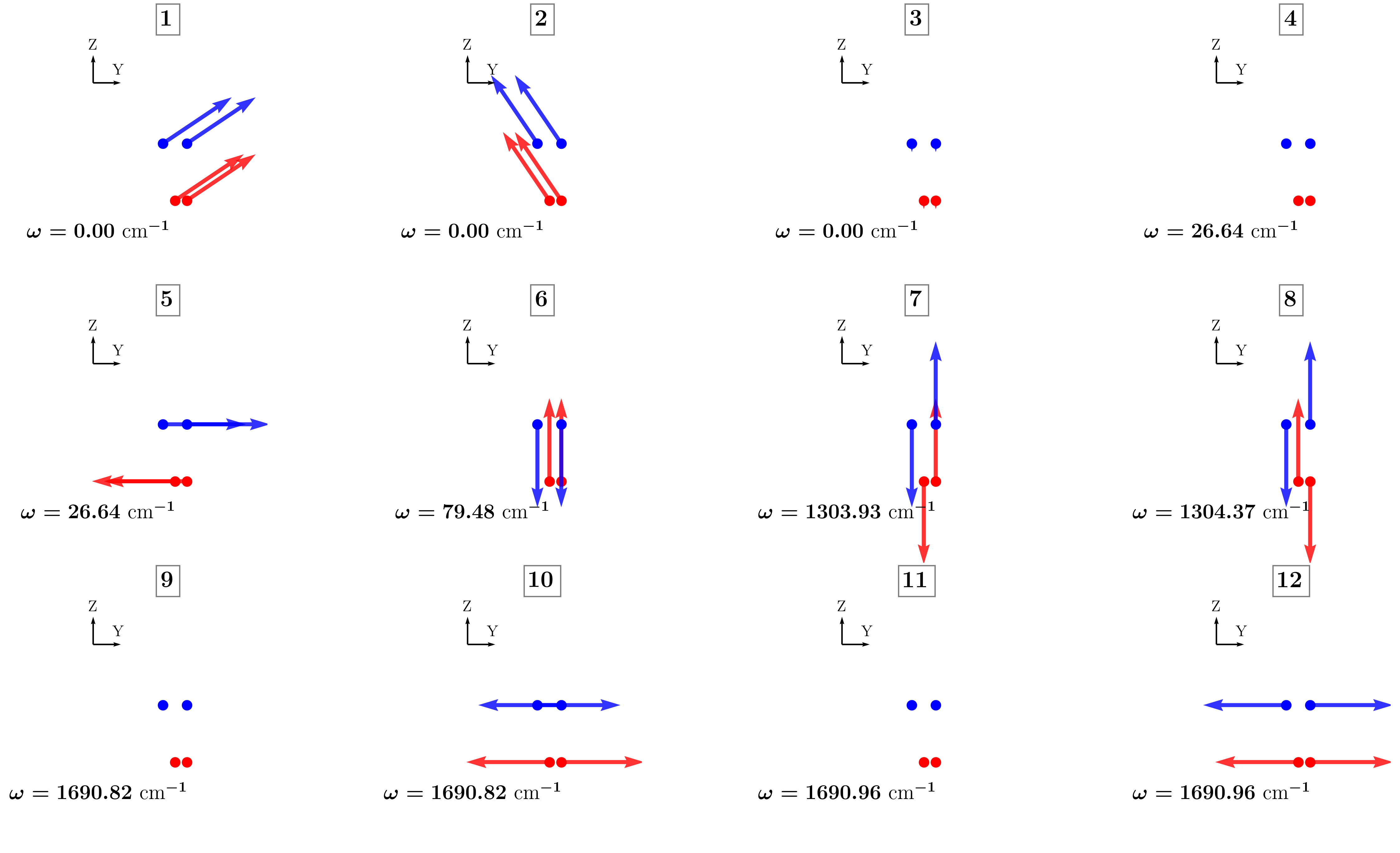}
    \caption{X-projections of the unit
    cell $\Gamma$ phonons in AB stacked bilayer graphene}
    \label{fig:ABphon-X}
\end{figure}

\begin{figure}
    \centering
    \includegraphics[width=\textwidth]{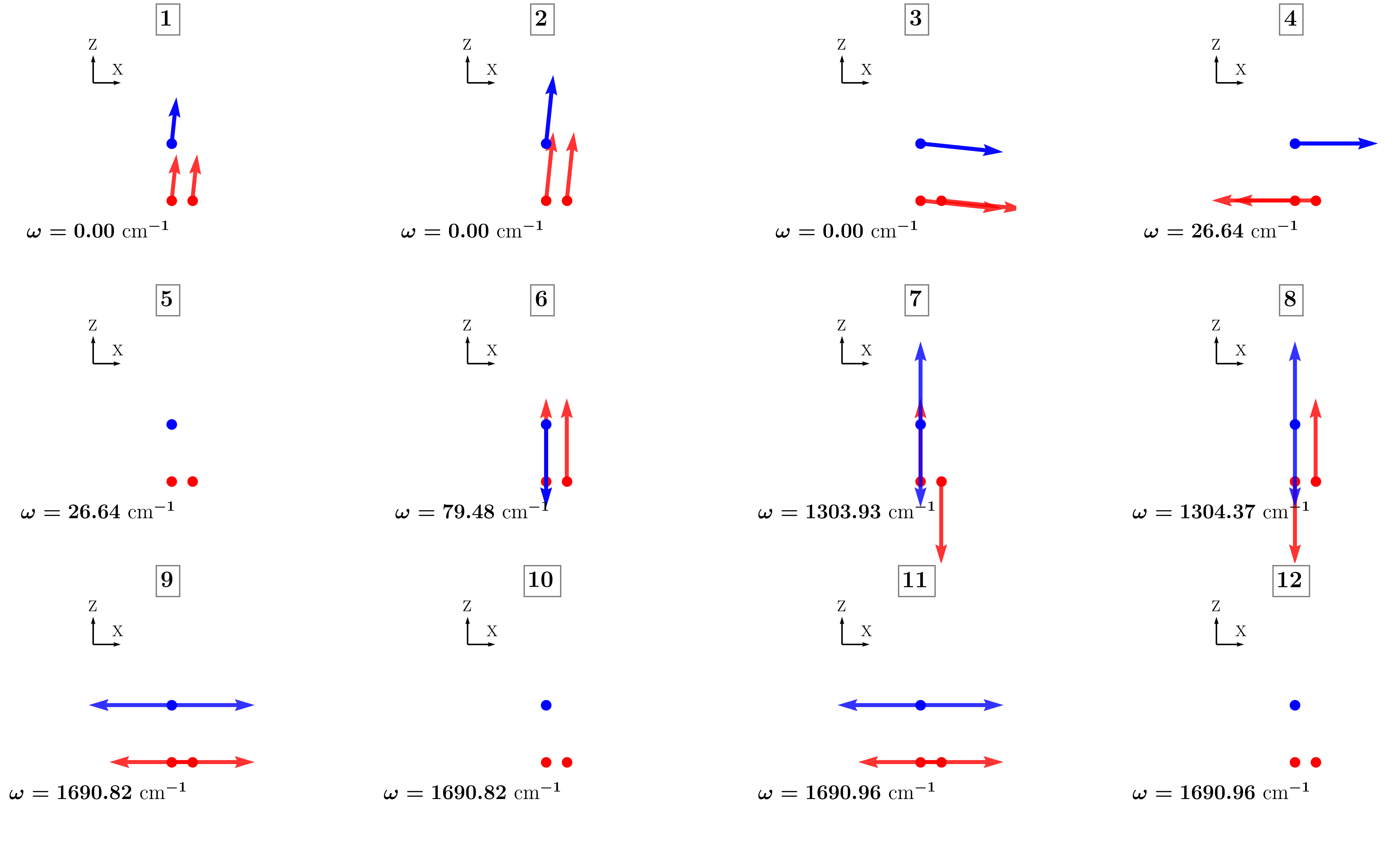}
    \caption{Y-projections of the unit
    cell $\Gamma$ phonons in AB stacked bilayer graphene}
    \label{fig:ABphon-Y}
\end{figure}

\begin{figure}
    \centering
    \includegraphics[width=\textwidth]{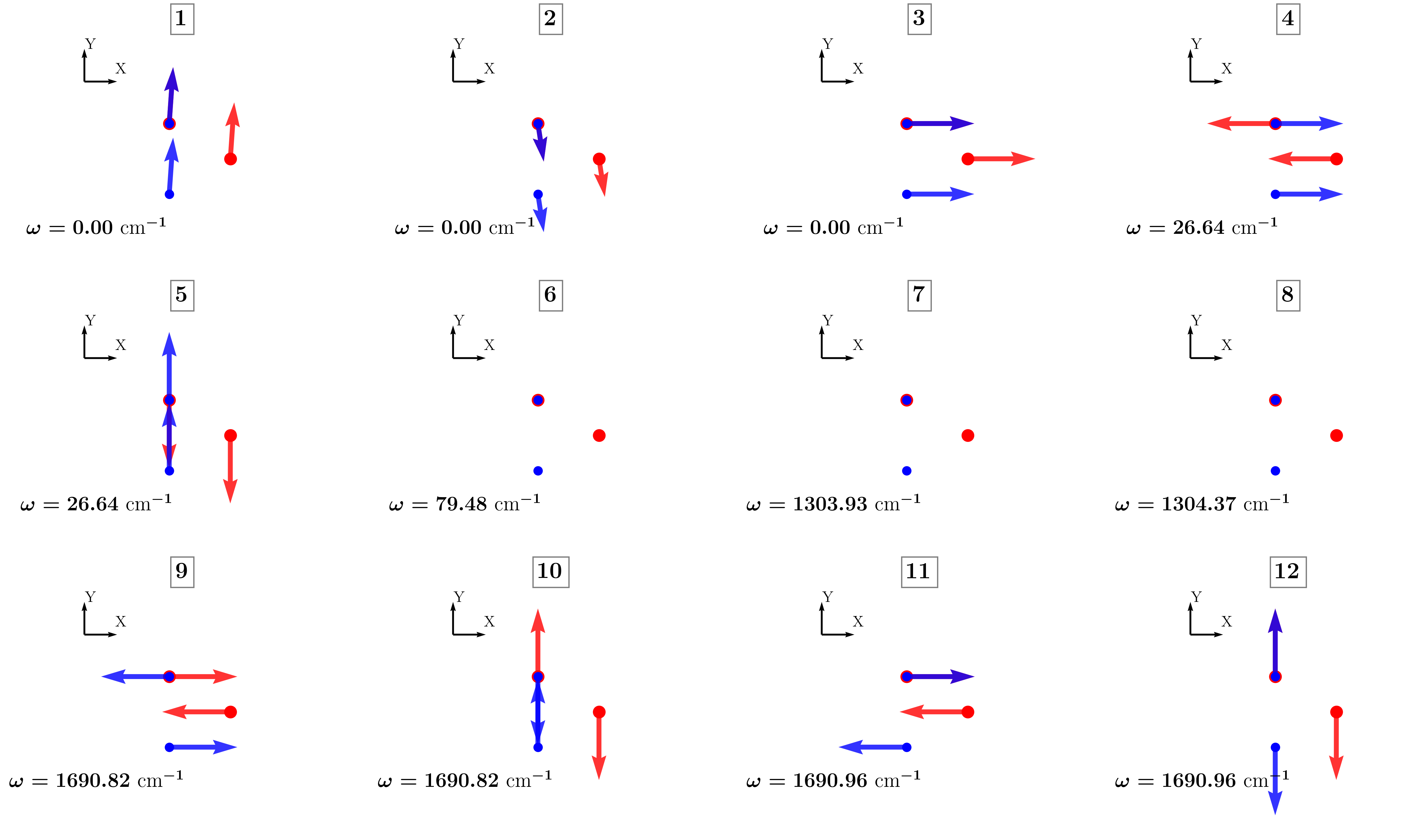}
    \caption{Z-projections of the unit
    cell $\Gamma$ phonons in AB stacked bilayer graphene}
    \label{fig:ABphon-Z}
\end{figure}

The bilayer modes were analyzed by plotting the eigenvectors at the $\Gamma$ point. The projections in the planes perpendicular to each Cartesian direction are shown in Fig \ref{fig:ABphon-X},\ref{fig:ABphon-Y} and \ref{fig:ABphon-Z}. The blue dots denote the atoms in the top layer while the red dots are the atoms in the bottom layer. The arrows show the displacement of the atoms in the direction of the eigenvector. From these projections, we can identify the E$_{\text{g}}$ modes in the bilayer system. The frequencies corresponding to each mode are also shown.

\newpage
\section{Electron-Phonon Calculations}
\subsection{Benchmarking}
\begin{figure}[h!]
    \centering
    \includegraphics[width=0.6\textwidth]{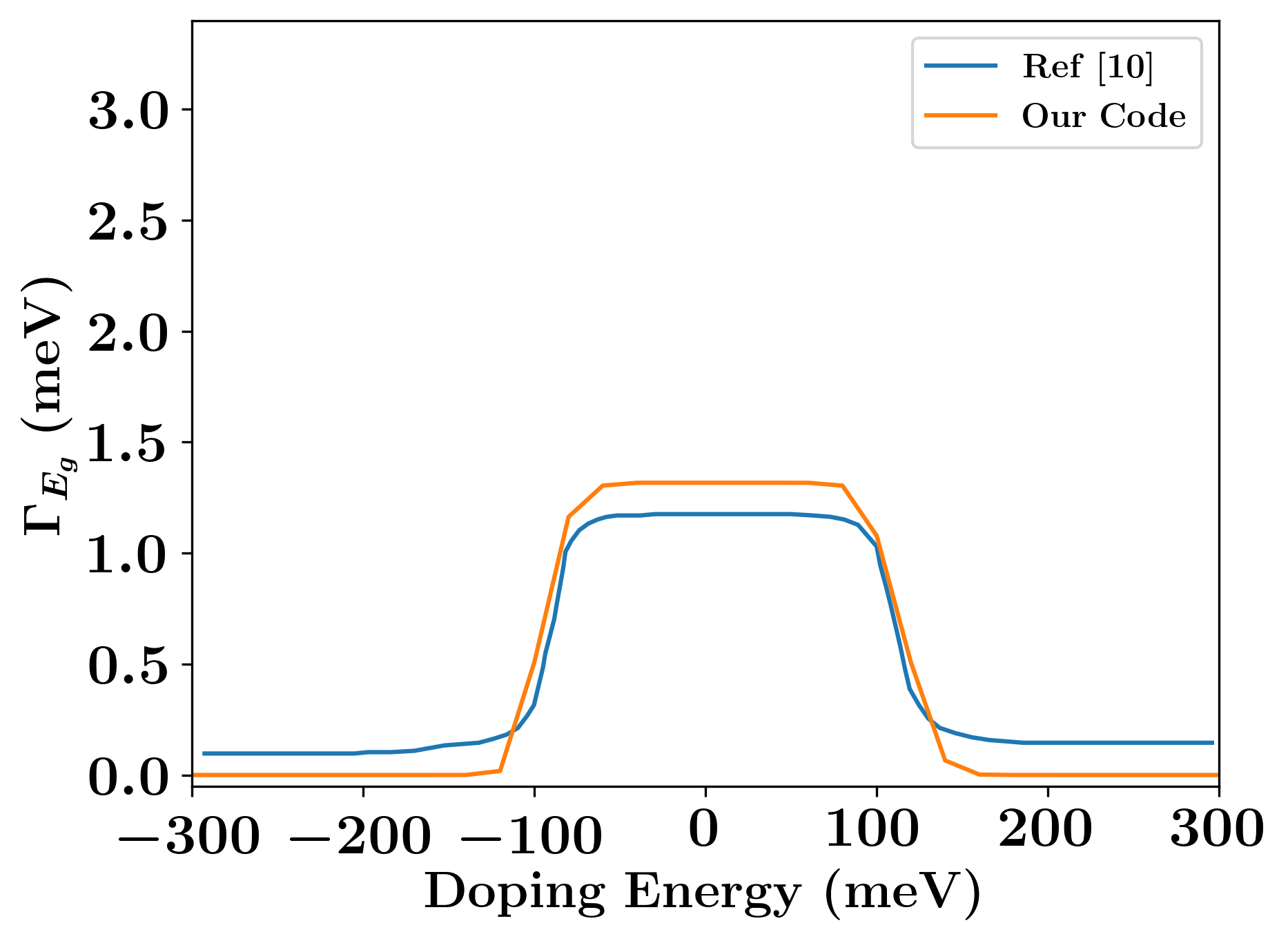}
    \caption{E$_{\text{g}}$ mode linewidth in AB bilayer from our approach compared with first principles results from \cite{park2008electron} }
    \label{fig:Benchmarking}
\end{figure}

\newpage
\section{Fits to the el-ph spectral functions}
Although the spectral functions plots from eqn.(8) of the main text may resemble Raman intensity plots, it is crucial to recognize that they are not equivalent.
While the positions of the Raman peaks and their associated linewidths are likely correctly extracted by fitting Lorentzians to the spectral function, the spectral function values themselves should not be misconstrued as corresponding to the measured Raman intensities. 
\begin{figure}[ht]
    \centering
    \includegraphics[width=0.975\textwidth]{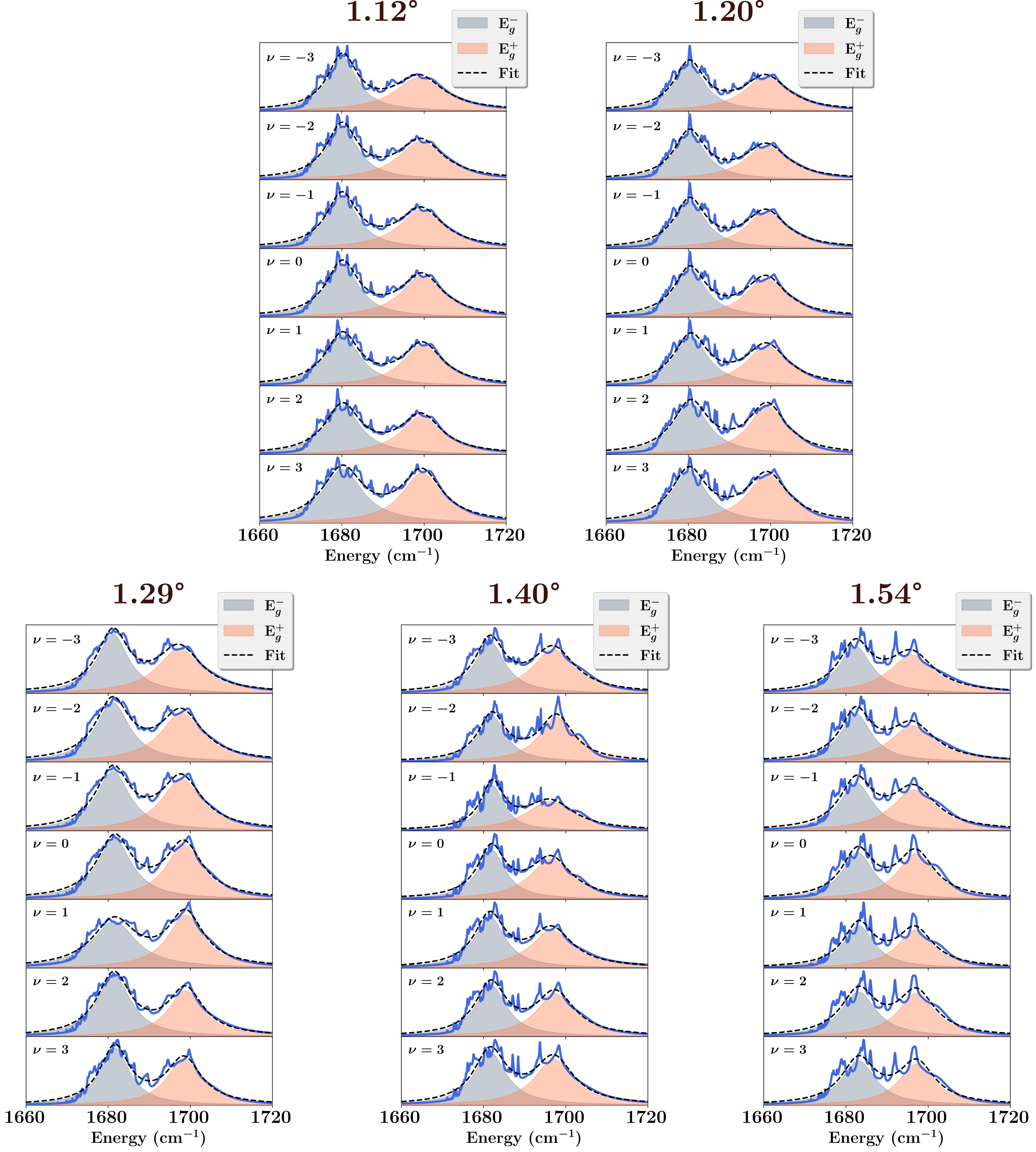}
    \caption{Lorentzian fits for the electron-phonon spectral function}
    \label{fig:elph_fits_total}
\end{figure}

\newpage
\section{Spatial resolution of the spectral function} 
\begin{figure}[h!]
    \centering
    \includegraphics[width=0.95\textwidth]{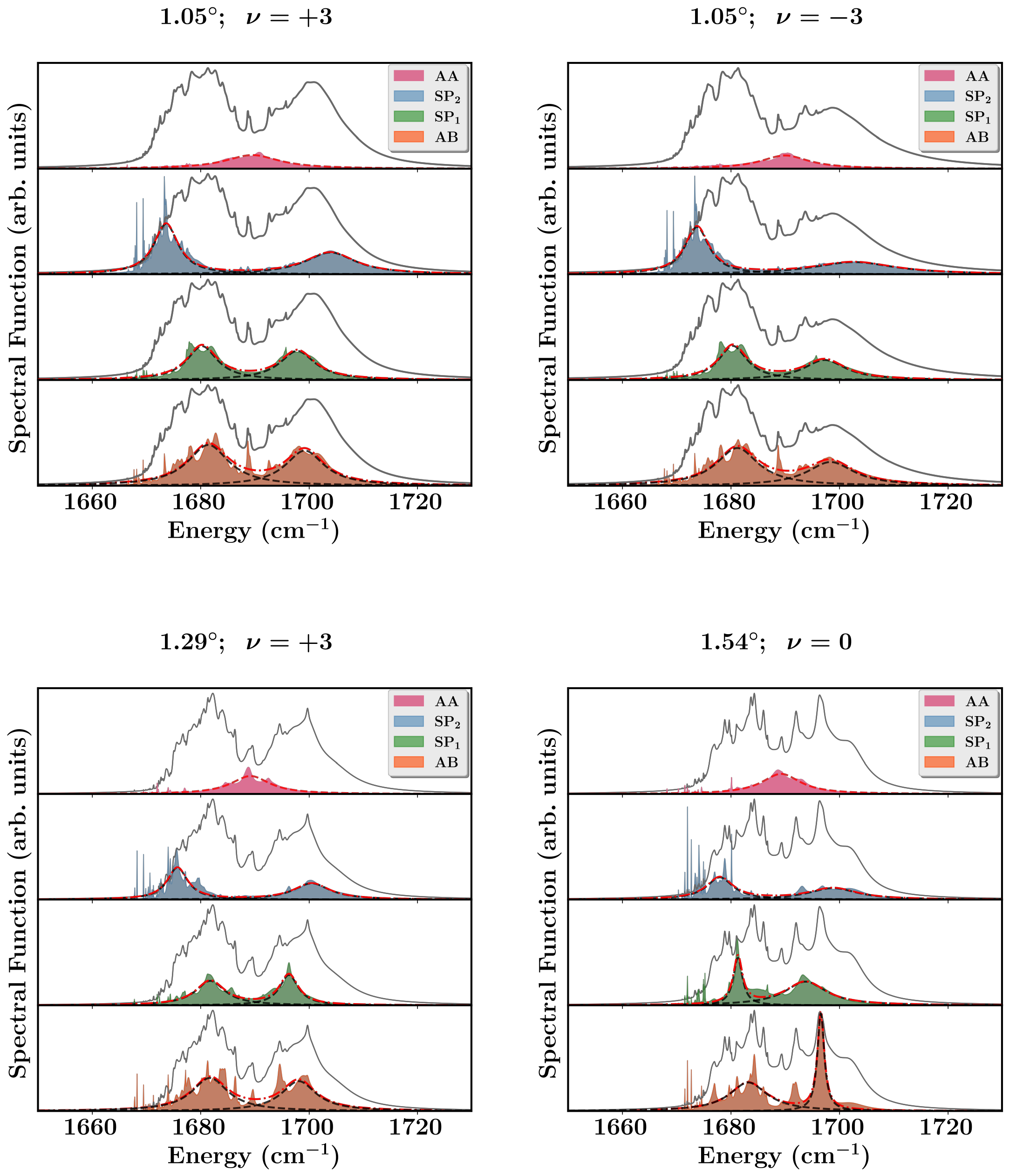}
    \caption{The projections onto different stacking regions are shown for some angles at different dopings. The total spectrum is marked in each of the panels}
    \label{fig:elph_fits_spatial}
\end{figure}

\newpage
\section{Joint Density of States}
\begin{figure}[ht]
    \centering
    \includegraphics[width=0.95\textwidth]{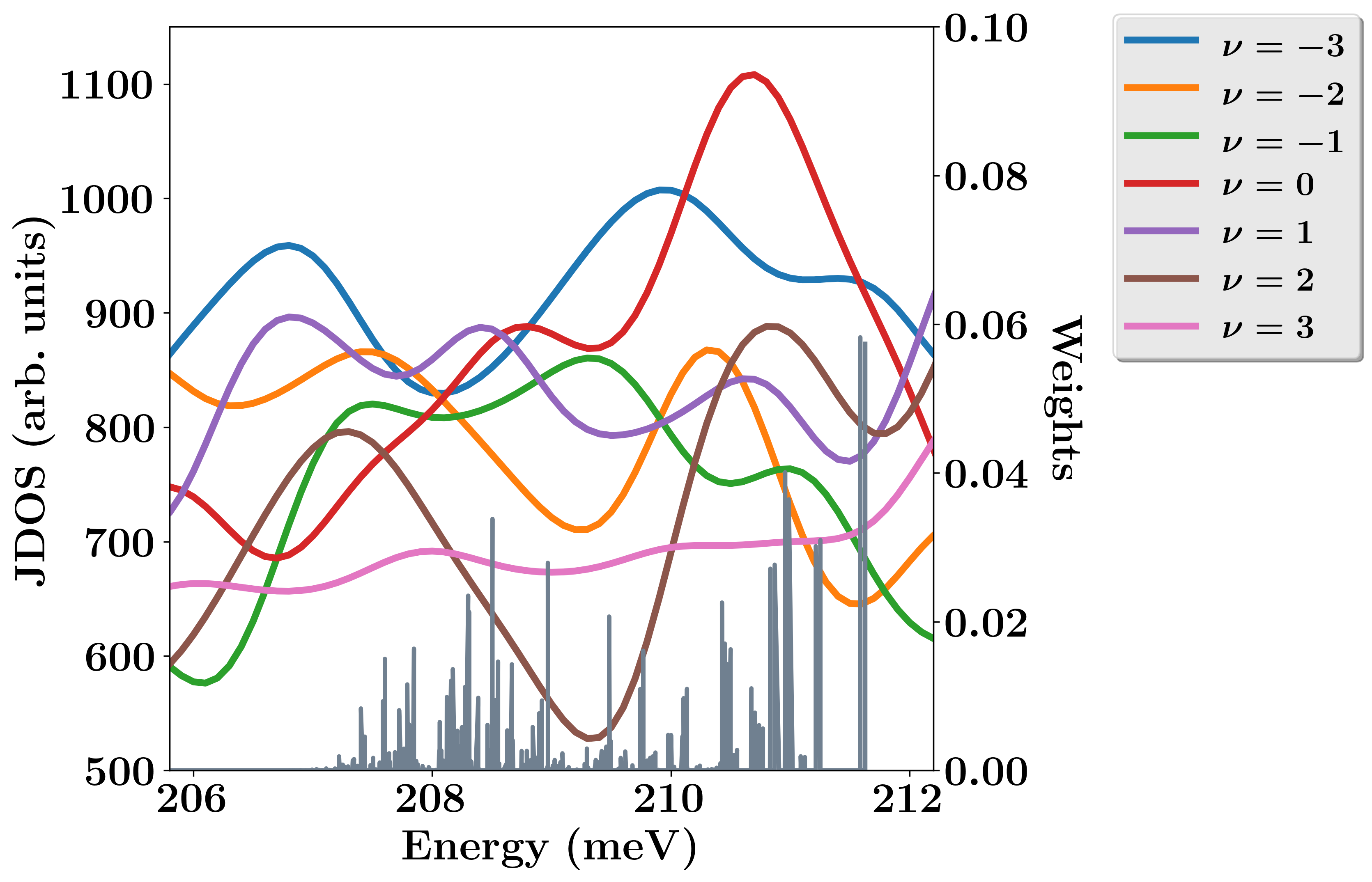}
    \caption{Joint Density of States $\big[J(\omega) = \sum_{m,n,\mathbf{k}} \ (f_{\mathbf{k}}^{n} - f_{\mathbf{k}}^{m}) \delta(\epsilon_{\mathbf{k}}^{m}-\epsilon_{\mathbf{k}}^{n}-\omega)\big]$ for $1.05^{\circ}$ TBG at different integer fillings. The projected weights of each of the moiré phonon modes onto the bilayer like E$_{\text{g}}$ mode is shown by the grey lines. The apparent anomaly in the abrupt increase in the linewidth of the E$_{\text{g}}^{+}$ mode (fig. 2a in the main manuscript) can be explained by looking at the position of the E$_{\text{g}}^{+}$ weights and the peaks in the joint density of states. The peak in the moiré mode projections corresponding to the E${_{\text{g}}^{+}}$ mode aligns with the peak in the joint density of states at $\nu = 0$ (the red curve). This causes an enhancement of the E${_{\text{g}}^{+}}$ linewidth for $1.05^{\circ}$ TBG at charge neutrality.}
    \label{fig:JDOS}
\end{figure}

\newpage
\section{Contribution of the Hartree Interactions to the el-ph coupling}
Within a localized orbital framework, the electron-phonon coupling matrix elements arise from all the changes in the Hamiltonian due to the atomic displacements corresponding to the phonon modes \cite{agapito2018ab}
\begin{equation}
    g_{\mathbf{k},\mathbf{q}}^{mns} =  l_{\mathbf{q}s}\sum_{\kappa\alpha}\Psi_{\mathbf{q}s, \kappa\alpha}^{\mathbf{M}}\sum_{pp';ij}
	e^{i\mathbf{k}\mathbf{R_p}} e^{-i(\mathbf{k+q})\mathbf{R_{p'}}} \phi_{\mathbf{k+q},mj}^{*} \phi_{\mathbf{k},ni}
    \langle  
    \Phi_{j};\mathbf{R}_{p'} \lvert
    \frac{\partial \hat{\mathbf{H}}}{\partial \tau_{0}^{\kappa,\alpha}} \rvert
    \Phi_{i};\mathbf{R}_p
    \rangle
\end{equation}
where $\lvert \Phi_{i}; \mathbf{R}_p \rangle$ denotes the orbital at $\tau_{pi}$ and $\tau_{pi}=\tau_{0i}+\mathbf{R}_p$. The coefficients of the electron orbitals in this basis are denoted by $\phi_{\mathbf{k},ni}=e^{-i\mathbf{k}\mathbf{R}_p}\langle \Phi_{i}; \mathbf{R}_p \lvert n\mathbf{k}\rangle$.  \\
Substituting the non-interacting tight binding model without any interactions, 
\begin{equation}
    \hat{\mathbf{H}} = \sum_{p\kappa;p'\kappa'} t(\tau_{p}^{\kappa} - \tau_{p'}^{\kappa'}) \lvert \Phi_{\kappa};\mathbf{R}_{p} \rangle \langle \Phi_{\kappa'};\mathbf{R}_{p'} \rvert
\end{equation}
we get the contribution from the transfer integral to the electron-phonon coupling matrix element following Refs. \cite{agapito2018ab,choi2018strong} and in the notations described in the main text as:  \begin{equation}
    g_{\mathbf{k},\mathbf{q}}^{mns} (\text{hop}) = l_{\mathbf{q}s}\sum_{\kappa\alpha}\Psi_{\mathbf{q}s, \kappa\alpha}^{\mathbf{M}}\sum_{pi}\frac{\partial}{\partial x_{\alpha}} t(\tau_{0}^{\kappa} - \tau_{p}^{i}) \\
	\left( e^{i\mathbf{k}\mathbf{R_p}} \phi_{\mathbf{k+q},m\kappa}^{*} \phi_{\mathbf{k},ni}  + 
	e^{-i(\mathbf{k+q})\mathbf{R_p}} \phi_{\mathbf{k+q},mi}^{*}\phi_{\mathbf{k},n\kappa}\right)
    \label{eqn:couple-hop}
\end{equation}
When the Hamiltonian has the additional onsite potential (arising from the electron-electron interactions) at each atomic site,
\begin{equation}
    \hat{\mathbf{H}} = \sum_{p\kappa;p'\kappa'} t(\tau_{p}^{\kappa} - \tau_{p'}^{\kappa'}) \lvert \Phi_{\kappa};\mathbf{R}_{p} \rangle \langle \Phi_{\kappa'};\mathbf{R}_{p'} \rvert + \sum_{p\kappa} \epsilon_{\kappa} \lvert \Phi_{\kappa};\mathbf{R}_{p} \rangle \langle \Phi_{\kappa};\mathbf{R}_{p} \rvert
\end{equation}
there is an additional contribution to the electron-phonon coupling arising from the onsite term:
\begin{align}
    g_{\mathbf{k},\mathbf{q}}^{mns} (\text{ons})=&  l_{\mathbf{q}s}\sum_{\kappa\alpha}\Psi_{\mathbf{q}s, \kappa\alpha}^{\mathbf{M}} \sum_{pp';ij} e^{-i(\mathbf{k+q})\mathbf{R}_p} e^{i\mathbf{k}\mathbf{R}_{p'}} 
    \phi_{\mathbf{k+q},mj}^{*} \phi_{\mathbf{k},ni} \nonumber 
    \sum_{\zeta,\beta}
    \langle  
    \Phi_{j};\mathbf{R}_{p'} \lvert
    \Phi_{\beta};\mathbf{R}_{\zeta} \rangle 
    \frac{\partial \epsilon_{\beta}}{\partial \tau_{0}^{\kappa,\alpha}} 
    \langle
    \Phi_{\beta};\mathbf{R}_{\zeta}
    \rvert
    \Phi_{i};\mathbf{R}_p
    \rangle \nonumber \\
    =& l_{\mathbf{q}s}\sum_{\kappa\alpha}\Psi_{\mathbf{q}s, \kappa\alpha}^{\mathbf{M}} \sum_{pp';ij} e^{-i(\mathbf{k+q})\mathbf{R}_p} e^{i\mathbf{k}\mathbf{R}_{p'}} 
    \phi_{\mathbf{k+q},mj}^{*} \phi_{\mathbf{k},ni} 
    \sum_{\zeta,\beta}
    \delta_{j,\beta} \delta_{p',\zeta}  \ 
    \frac{\partial \epsilon_{\beta}}{\partial \tau_{0}^{\kappa,\alpha}} \ 
    \delta_{i,\beta}\delta_{p,\zeta} \nonumber \\
    =& l_{\mathbf{q}s}\sum_{\kappa\alpha}\Psi_{\mathbf{q}s, \kappa\alpha}^{\mathbf{M}} \sum_{pp';ij} e^{-i(\mathbf{k+q})\mathbf{R}_p} e^{i\mathbf{k}\mathbf{R}_{p'}} 
    \phi_{\mathbf{k+q},mj}^{*} \phi_{\mathbf{k},ni}
    \frac{\partial \epsilon_{j}}{\partial \tau_{0}^{\kappa,\alpha}} \delta_{i,j} \delta_{p,p'} \nonumber \\
    =& l_{\mathbf{q}s}\sum_{\kappa\alpha}\Psi_{\mathbf{q}s, \kappa\alpha}^{\mathbf{M}}\sum_{p,i} e^{-i\mathbf{q}\mathbf{R}_p} \phi_{\mathbf{k+q},mi}^{*} \phi_{\mathbf{k},ni} 
    \frac{\partial \epsilon_{i}}{\partial \tau_{0}^{\kappa,\alpha}}
     \label{eqn:coupling-onsite}
\end{align}
In our work the onsite potential is described by the form proposed by Zachary et. al \cite{goodwin2020hartree}.
\begin{align}
    \epsilon_{i} &= \sum_{j}(\mathbf{n}_j - \bar{\mathbf{n}}) \sum_{\mathbf{R_p}} \text{W}_{\mathbf{R_{p}},ij} \nonumber \\
    \text{W}_{\mathbf{R_{p}},ij} &= \frac{e^2}{4\pi\epsilon_0\epsilon_{\text{bg}}}
    \sum_{m=-\infty}^{\infty}\frac{(-1)^m}{\sqrt{(\mathbf{R_p} + \tau_{0}^{j} - \tau_{0}^{i})^2 + (2m\xi)^2}} 
\end{align}
where $\mathbf{n}_j$ is the number density at each atomic site $j$ and $\bar{\mathbf{n}}$ is the average of $\mathbf{n}_j$. At a filling $\nu$, the average of the number density, $\bar{\mathbf{n}} = 1+\nu/N$ where $N$ is the total number of atoms in the moiré unit cell. Here we have assumed that the TBG is encapsulated by a dielectric substrate of thickness $\xi$ and dielectric constant $\epsilon_{\text{bg}}$. The values of these parameters were obtained from Ref. \cite{goodwin2020hartree}. 

\begin{figure}
    \centering
    \includegraphics[width=0.75\textwidth]{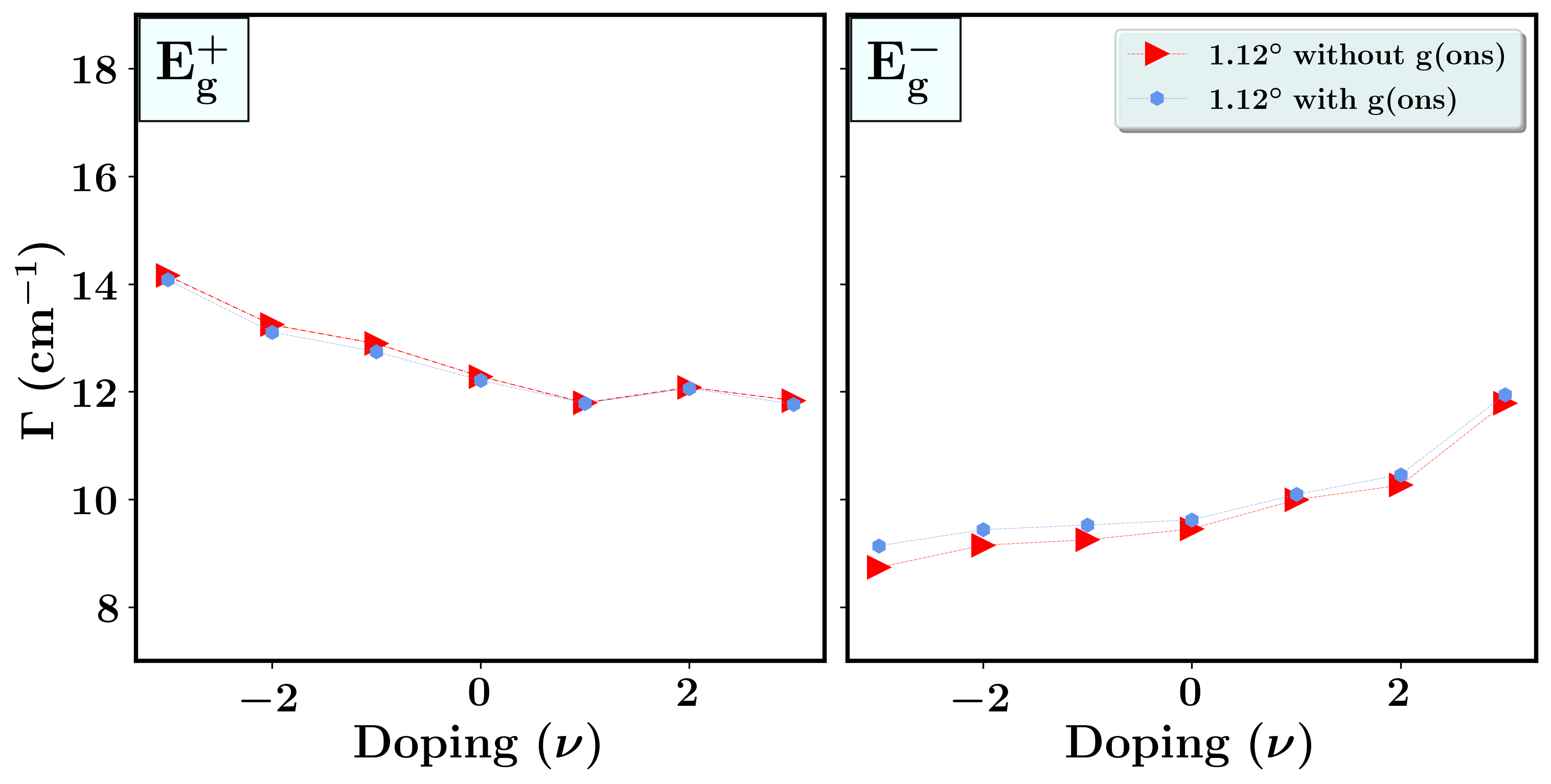}
    \caption{E$_{\text{g}}$ mode linewidths at $1.12^{\circ}$ twist with the correction due to $g_{\mathbf{k,q}}^{mns}(\text{ons})$ included.}
    \label{fig:with_g_ons}
\end{figure}

\noindent The derivative of the onsite potential in eqn.(\ref{eqn:coupling-onsite}) can hence be evaluated as 
\begin{align}
    \frac{\partial \epsilon_{i}}{\partial \tau_{0}^{\kappa,\alpha}} &=  \Big[ \sum_{j}\frac{\partial(\mathbf{n}_j - \bar{\mathbf{n}})}{\partial \tau_{0}^{\kappa,\alpha}} \sum_{p} \text{W}_{\mathbf{R_{p}},ij} \Big] + 
    \Big[
    \sum_{j}(\mathbf{n}_j - \bar{\mathbf{n}}) \sum_{p} \frac{\partial\text{W}_{\mathbf{R_{p}},ij}}{\partial \tau_{0}^{\kappa,\alpha}} \Big]
\end{align}
The change in the number density $\frac{\partial(\mathbf{n}_j - \bar{\mathbf{n}})}{\partial \tau_{0}^{\kappa,\alpha}}$ can be safely neglected as this is of higher order in the displacements. Consequently we get 
\begin{equation}
    \frac{\partial \epsilon_{i}}{\partial \tau_{0}^{\kappa,\alpha}} =  
    \sum_{j}(\mathbf{n}_j - \bar{\mathbf{n}}) \sum_{p} \frac{\partial\text{W}_{\mathbf{R_{p}},ij}}{\partial \tau_{0}^{\kappa,\alpha}}
\end{equation}
The total electron phonon coupling matrix element is obtained as 
\begin{equation}
    g_{\mathbf{k},\mathbf{q}}^{mns} = g_{\mathbf{k},\mathbf{q}}^{mns} (\text{hop}) + g_{\mathbf{k},\mathbf{q}}^{mns} (\text{ons})
\end{equation}
Within our framework, the addition of the onsite contribution does not result in a significant change to the linewidths obtained, is clear from fig. \ref{fig:with_g_ons}. 
So it can be concluded that $g_{\mathbf{k,q}}^{mns}\text{(hop)}$ is the dominant term.

\newpage
\section{Anharmonic Contribution}

The anharmonic contribution to the linewidth is computed from the power spectra obtained via the fourier transform of the mode projected velocity autocorrelation function (MVACF)	\cite{maity2020phonons}
    \begin{align}
            \zeta_{\mathbf{q}s}(\omega) = \int_{0}^{\infty} dt \ \big\langle \mathbf{V}_{\mathbf{q}s}(0) \mathbf{V}_{\mathbf{q}s}^{*}(t)\big\rangle e^{-i\omega t}
	\end{align}
The MVACF is defined as
\begin{equation}
    \big\langle \mathbf{V}_{\mathbf{q}s}(0) \mathbf{V}_{\mathbf{q}s}^{*}(t)\big\rangle = \lim_{\tau \to \infty} \frac{1}{\tau} \int_{0}^{\tau} \mathbf{V}_{\mathbf{q}s}(t') \centerdot \mathbf{V}_{\mathbf{q}s}^{*}(t+t') \text{d}t'
\end{equation}
where the $\mathbf{V}_{\mathbf{q}s}(t)$ is the mode projected velocity defined by 
\begin{align}
	\begin{split}
        \mathbf{V}_{\mathbf{q}s}(t) &= \tilde{\mathbf{v}}_{\mathbf{q}}(t) \cdot \hat{\eta}_{\mathbf{q}s} \\
		\tilde{\mathbf{v}}_{\mathbf{q}}^{\mu}(t) &= \sqrt{m_{\mu}} \sum_{j}  e^{-i\mathbf{q}\cdot\mathbf{r}_{j}(t)}\mathbf{v}_{j}(t) \ ; \ \ \ \ 
		j \in \{\text{Atoms of type }\mu\}
	\end{split}
\end{align}
where  $\mathbf{v}_{i}(t)$ is the velocity at time $t$ for the particle $i$, $m_{\mu}$ is the mass of the atom of type $\mu$ and $\hat{\eta}_{\mathbf{q}s}$ is the polarization of the unit cell phonon mode $(\mathbf{q}s)$. \\
In our calculations since we are interested in only the $\mathbf{q} = \mathbf{0}$ phonons, the equations simplify to 
\begin{align}
    \begin{split}
        \tilde{\mathbf{v}}_{\mathbf{0}}^{\mu}(t) &= \sqrt{m_{\mu}} \sum_{j}  \mathbf{v}_{j}(t) \ ; \ \ \ \ 
		j \in \{\text{Atoms of type }\mu\} \\
        \mathbf{V}_{\mathbf{0}s}(t) &= \tilde{\mathbf{v}}_{\mathbf{0}}(t) \cdot \hat{\eta}_{\mathbf{0}s} \\
        \big\langle \mathbf{V}_{\mathbf{0}s}(0) \mathbf{V}_{\mathbf{0}s}^{*}(t)\big\rangle &= \lim_{\tau \to \infty} \frac{1}{\tau} \int_{0}^{\tau} \mathbf{V}_{\mathbf{0}s}(t') \centerdot \mathbf{V}_{\mathbf{0}s}^{*}(t+t') \text{d}t' \\
        \zeta_{\mathbf{0}s}(\omega) &= \int_{0}^{\infty} dt \ \big\langle \mathbf{V}_{\mathbf{0}s}(0) \mathbf{V}_{\mathbf{0}s}^{*}(t)\big\rangle e^{-i\omega t}
    \end{split}
\end{align}

\noindent $\Gamma_{\text{ph-ph}}$ is obtained as the FWHM of the Lorentzian fitted to the G mode projected power spectra. 
The calculations were performed using a system containing $12 \times 12$ moiré unit cells for the angles $1.40$ and $1.54$, and $10 \times 10$ moiré unit cells for the other angles. The systems were equilibrated at the final temperature under the NVE ensemble for $100$ ps. The velocities of all the atoms were collected every $4$ fs for $800$ ps in the NPT ensemble.

\begin{figure}
    \centering
    \includegraphics[width=\textwidth]{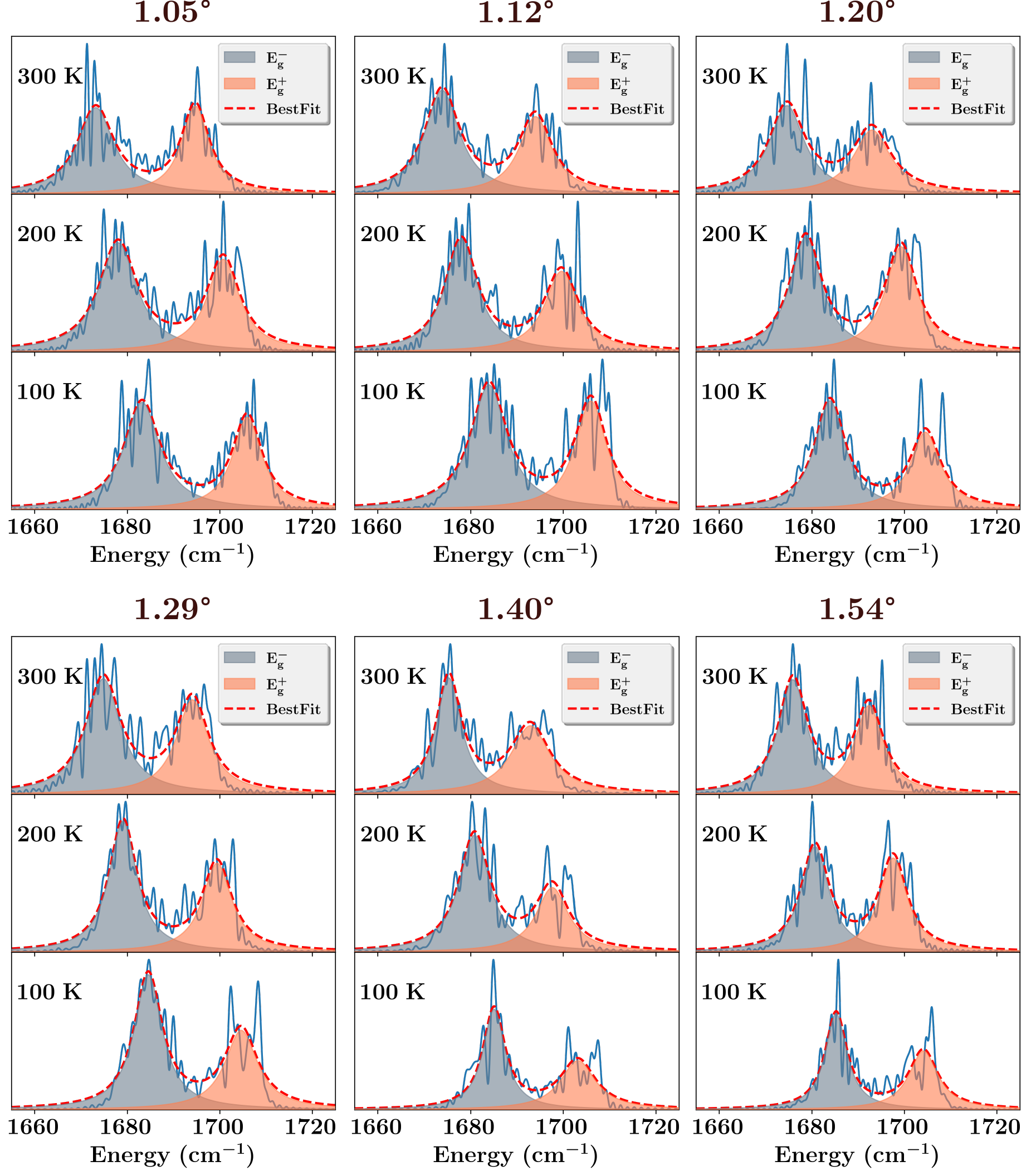}
    \caption{The fits for the mode projected power spectra at various angles of twist, for different temperatures. The FWHM of these fits give the linewidth while the position of the Lorentzians provide the lineshifts.}
    \label{fig:anharm_fits}
\end{figure}

\noindent The fits at the different angles of twist we have considered in our calculations are shown in Fig. \ref{fig:anharm_fits}.

\newpage
\bibliography{bibtex.bib}